\documentclass{svproc}
\usepackage{url}

\usepackage{graphicx}
\usepackage{multicol}
\usepackage{footmisc}

\begin{document}
\mainmatter              
\title{Role of fluctuations on the pairing properties of nuclei in the 
random spacing model}
\titlerunning{Random Spacing Model}  
%
\author{M. A. A. Mamun\inst{1}, C. Constantinou\inst{2} \and M. Prakash\inst{3}}
\authorrunning{M. A. A. Mamun et al.} 
%
\tocauthor{C. Constantinou, M. Prakash}
\institute{
Department of Physics and Astronomy, Ohio University, Athens, Ohio 45701, USA\\
\email{ma676013@ohio.edu}
\and
Department of Physics, Kent State University, Kent, OH 44242, USA\\
\email{cconsta5@kent.edu},
\and
Department of Physics and Astronomy, Ohio University, Athens, Ohio 45701, USA\\
\email{prakash@ohio.edu}
}

\maketitle              

\begin{abstract}
The influence of thermal fluctuations on fermion pairing
is investigated using a 
semiclassical treatment of fluctuations. When the average pairing gaps 
along with those differing by one standard deviation are used, the 
characteristic discontinuity of the specific heat at the critical temperature $T_c$ in the BCS 
formalism with its most probable gap becomes smooth. This indicates the 
suppression of a second order phase transition as experimentally 
observed in nano-particles and several nuclei.  Illustrative calculations  using 
the constant spacing model and the recently introduced random spacing model are presented. 

\keywords{Pairing Fluctuations, Superfluidity and Superconductivity, Random Spacing Model}
\end{abstract}

\section{Pairing in Systems of Large and Small Numbers of Particles}
The Bardeen-Cooper-Schriffer (BCS) theory of superconductivity [1,2] predicts a sharp
discontinuity in the constant-volume specific heat $C_V$ at a certain critical temperature
$T_c$ for which the pairing gap $\Delta$ of fermions determined from the  
gap and number equations [3,4]
\begin{eqnarray}
\label{gapeq}
{\cal G} &\equiv& \sum_k \frac {1}{E_k} \tanh \left(\frac{E_k}{2T} \right) - \frac 2G  \\
N &=& 
\sum_k  \left[ 1 - \frac{\epsilon_{k} - \lambda}{E_{k}} \tanh \left(\frac{E_{k}}{2T}\right) \right] + \frac {\Delta} {T} \frac {\partial \Delta}{\partial \alpha}~{\cal G} 
\label{neq}
\end{eqnarray}
vanishes. 
Above, $G$ is the strength of the pairing interaction, $T$ is the temperature, $\lambda$ is the chemical potential, 
$\alpha=\lambda/T$, $E_k={\sqrt{(\epsilon_k-\lambda)^2 + \Delta^2}}$ is the quasiparticle energy, and 
$\epsilon_k$ are the single particle (sp) energies.  The corresponding energy $E$ and entropy $S$ 
are obtained from 
\begin{eqnarray}
\label{Efluc}
E &=& 
\sum_k \epsilon_{k} \left[ 1 - \frac{\epsilon_{k} - \lambda}{E_{k}} \tanh \left(\frac{E_{k}}{2T}\right) \right] 
 - \frac{\Delta^{2}}{G} 
 - \left(\Delta^2  - \Delta T \frac {\partial \Delta}{\partial T} \right) {\cal G} \\
S &=&2 \sum_k \left\{\ln \left[1 + \exp \left(-\frac {E_k}{T}\right)\right] + 2~ \frac{E_k/T}{1 + \exp(E_k/T)} \right\} \nonumber \\
 &-& \frac {\Delta}{T} \left( \frac {\lambda}{T} \frac {\partial \Delta}{\partial \alpha} - T \frac {\partial \Delta}{\partial T} \right)  {\cal G} \,.
 \label{Sfluc}
\end{eqnarray}
These relations enable the evaluation of   $C_V = dE/dT|_{V,N} = T(\partial S/\partial T)|_{V,N}$. 

In a mean field description of the BCS theory, $(\partial \Omega/\partial \Delta)|_T = 0={\cal G}$ which leads to the most probable gap $\Delta_{\rm mp}$.
 For systems with large numbers of particles, fluctuations in the order parameter $\Delta$ are very small as the probability distribution  %
\begin{eqnarray}
\label{PDelta}
 P(\Delta) &\propto& \exp[ -\Omega(T,\Delta) / T ] \,, \\
\Omega(T,\Delta) &=& \sum_k (\epsilon_k - \lambda - E_k) - 2T \sum_k \ln \left[ 1 + \exp \left(-\frac {E_k}{T} \right)  \right]  
+ \frac {\Delta^2}{G} \,,
\label{Omega}
\end{eqnarray}
where $\Omega$ is the grand potential, is very sharply peaked at $\Delta_{\rm mp}$.   In this case, Eqs. (\ref{neq})-(\ref{Sfluc}) revert back to the standard mean field BCS equations. 
For $\Delta \neq \Delta_{\rm mp}$, ${\cal G}\neq 0$,  and Eqs. (\ref{neq})-(\ref{Sfluc}) and hence $C_V$ receive additional contributions.

\begin{figure}
\includegraphics[width=1.0\columnwidth]{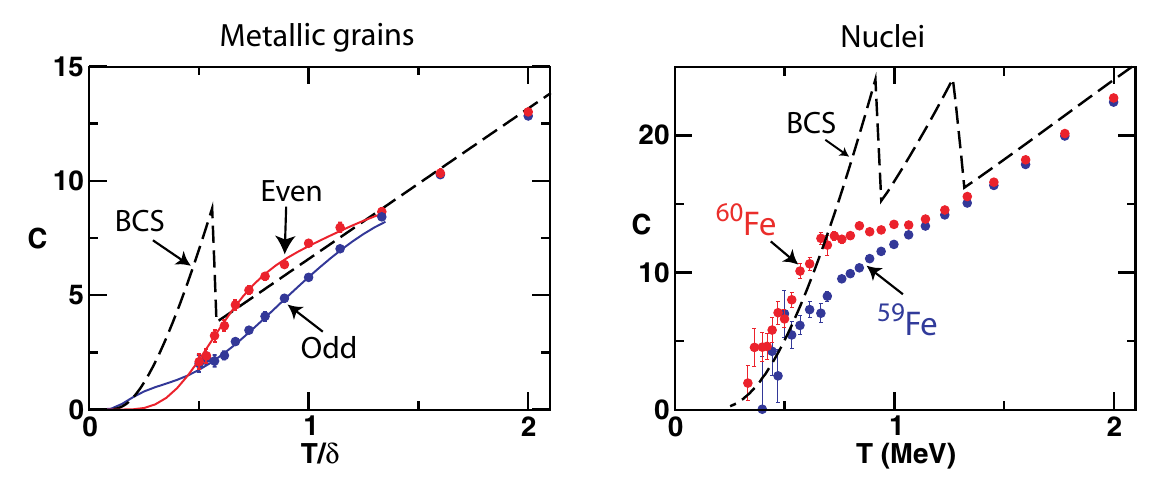}
\caption{Specific heat in gold nano particles (left) and  iron isotopes (right) demonstrating the 
disappearance of a second order phase transition present in the mean field BCS formalism. Figure adapted from Alhassid [6]. }
\label{Cvrnp197}
\end{figure}

In systems with small numbers of particles, fluctuations in $\Delta$ are not small. 
As first noted by Anderson in his paper ``Theory of Dirty Superconductors'' [5],  the pairing phenomenon is suppressed due to large fluctuations in $\Delta$ which in turn leads to a ``shoulder-like'' or ``S-shaped'' smooth curve for $C_V$ vs $T$. That a similar suppression would occur in nuclei also was first noted by Moretto in Ref. [3]. 
Experimental verifications of the absence of a sharp second order phase transition due to pairing in nanoparticles and nuclei 
are shown in Fig.  \ref{Cvrnp197}, which also contains results of Auxiliary Field Monte Carlo (AFMC) calculations  for $C_V$ vs $T$ including fluctuations by Alhassid et al [6]. 

\section{Fluctuations in the Order Parameter $\Delta$}

Fluctuations can arise from many sources. When $T$ is too low or $\Delta$ varies too rapidly with time, a thermodynamic treatment becomes inadequate and a fully quantum approach that accounts for correlations beyond mean field theory, pairing vibrations and suppression of pairing due to rotational motion, etc., becomes necessary [7-17]. A semiclassical treatment of thermal fluctuations based on Eq. (\ref{PDelta}) and ${\cal G}\neq 0$ in Eqs. (\ref{gapeq})-(\ref{Sfluc}) 
is afforded when $\Delta$ is strongly coupled to all other intrinsic degrees of freedom, that is 
when $\Delta >> \delta$,  where $\delta=1/g$ is the mean level spacing  of the sp  energy levels near the Fermi sea  [3,7].
For infinite systems 
(e.g., bulk nuclear matter)  $P(\Delta)$ approaches 
a delta function,  $\delta << \Delta$, whence fluctuations are negligible and mean field BCS   with ${\cal G}= 0$ is a 
reasonable description. In contrast, for small systems such as nanoparticles or light-to-medium heavy nuclei, 
$\delta \sim \Delta$ or  $\delta \geq \Delta$ particularly at $T\neq 0$,  fluctuations in $\Delta$ are large and suppress superconductivity and superfluidity. In this case, the mean field BCS approach  is no longer applicable as it neglects 
the influence of fluctuations. 

\begin{figure}
\includegraphics[width=0.5\columnwidth]{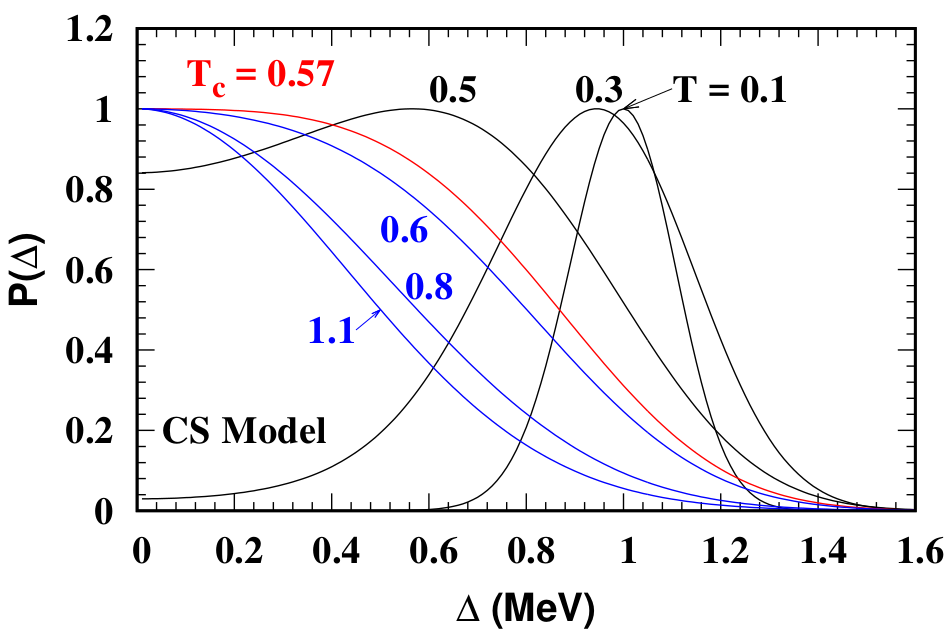}
\includegraphics[width=0.5\columnwidth]{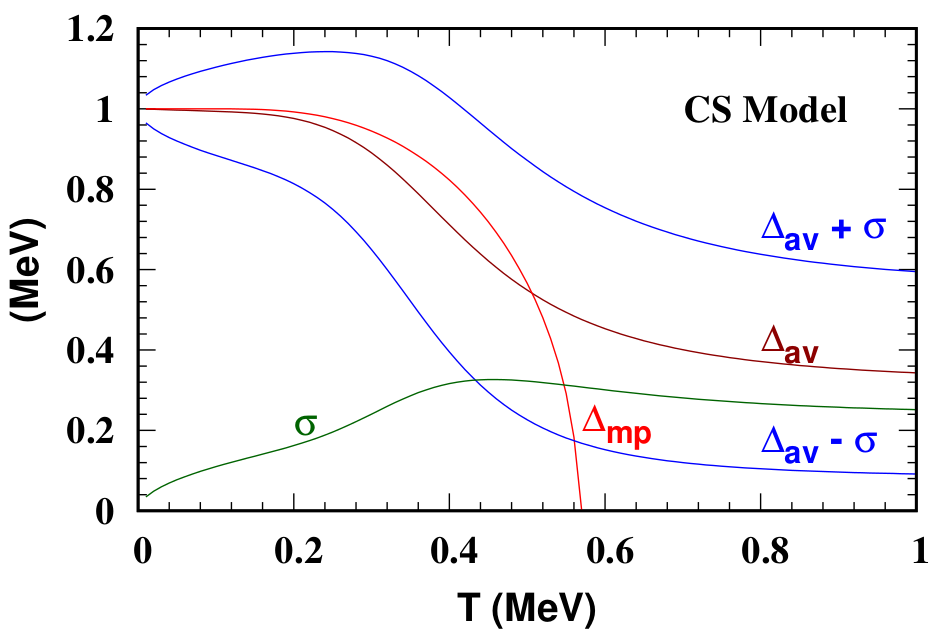}
\caption{Probability distribution of the pairing gap (left) and the gap as a function of 
temperature (right) in the constant spacing (CS) model. Figure adapted from [18]. }
\label{probdist}
\end{figure}

Figure \ref{probdist} illustrates the role of fluctuations in the constant spacing (CS) model with
 $g=5~{\rm MeV}^{-1}$ for doubly degenerate sp energy levels for $N=144$ and $\Delta(0)=1$ MeV at $T=0$. 
For this choice, $G=0.0581$ MeV, $\hbar\omega \simeq 41 N^{-1/3} = 7.78$ MeV, with levels uniformly distributed between $\pm 2\hbar\omega$ around $\lambda_{\rm mp}(0) = -1.3471$ MeV at $T=0$. The probability $P(\Delta)$ is normalized such that 
$P(\Delta_{\rm mp})=1$ for all $T$.   For all curves shown, $\lambda(T)$ vs $T$ is calculated for each $\Delta\neq\Delta_{\rm mp}$ using Eq. (\ref{neq})  thus ensuring number conservation.
The results in this figure are similar to those of Ref. [3] where $g=7~{\rm MeV}^{-1}$ was used. 

\begin{figure}
\includegraphics[width=0.5\columnwidth]{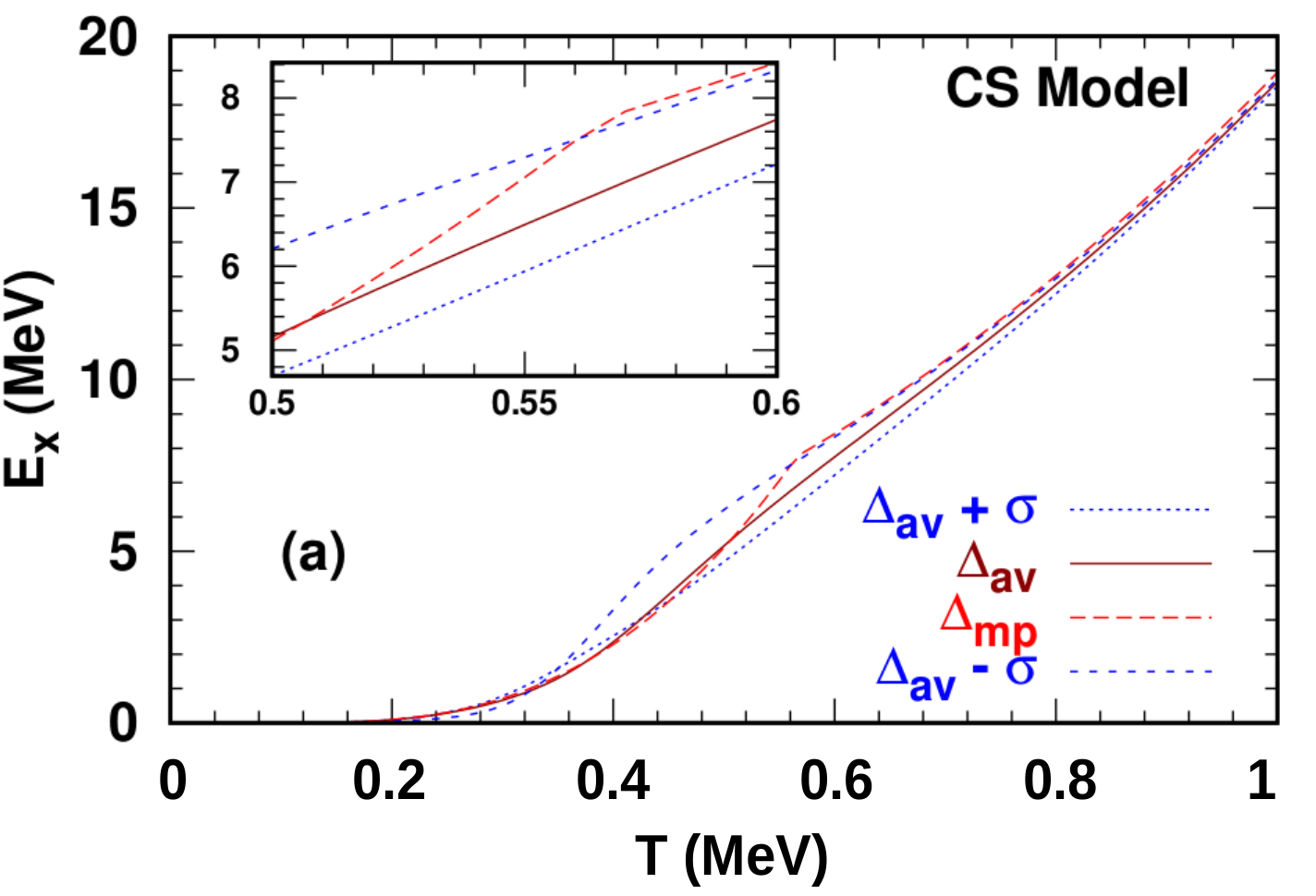}
\includegraphics[width=0.5\columnwidth]{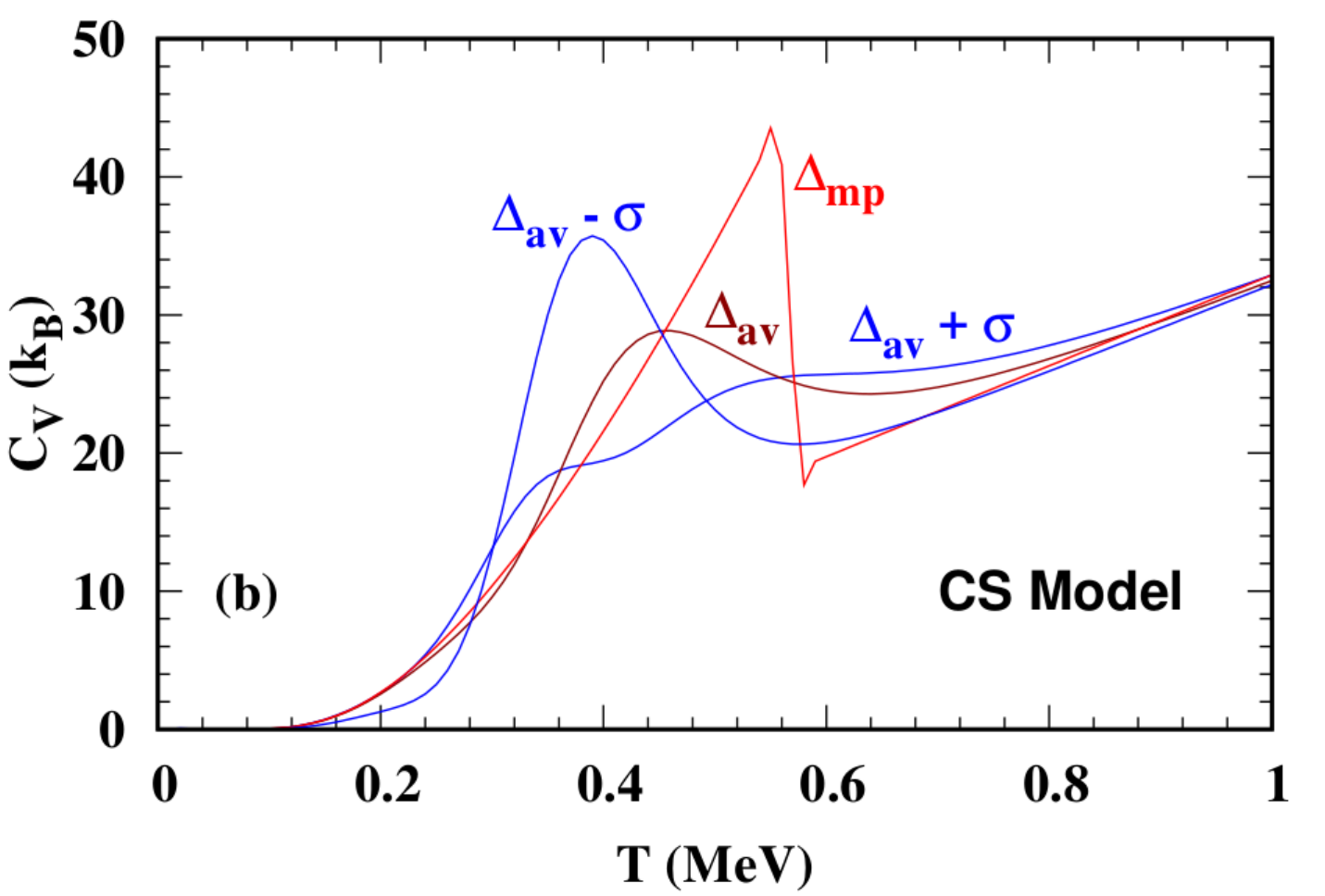}
\caption{Excitation energies (left) and specific heats at constant volume (right) with the gaps shown in Fig. \ref{probdist}. Figure adapted from [18]. }
\label{CSExCv}
\end{figure}

The salient features in the left panel of  Fig.  \ref{probdist} are (i)  for $T\simeq 0$,  $P(\Delta)$  is symmetrical around $\Delta_{{\rm mp}}$, (ii) with increasing $T$, $P(\Delta)$ becomes increasingly asymmetrical, and (iii) for $T\geq T_c\simeq 0.57$ MeV, $P(\Delta)$ is peaked at $\Delta=0$.  For all $T \neq 0$, the term involving the nonzero ${\cal G}$ in Eq. (\ref{neq}) gives significant contributions. As 
$P(\Delta)$ is very broad for $T\rightarrow T_c$ and beyond, use of  average thermodynamic quantities 
$<\tilde O> = {\sum \tilde O P(\Delta)}/{\sum P(\Delta)}$ is more appropriate than those 
with $\Delta_{\rm mp}$. The right panel of Fig.  \ref{probdist} provides contrasts between $\Delta_{\rm mp}$  and $\Delta_{\rm av}$ as well as for gaps differing by $\pm 1\sigma$  from   $\Delta_{\rm av}$. The latter gaps are nonzero  for $T>T_c$, unlike $\Delta_{\rm mp}$, indicating that pairing correlations persist beyond $T_c$. 
The  excitation energies $E_x = E(T) - E(0)$ and $C_V$ with the gaps shown in Fig. \ref{probdist} are shown in Fig. \ref{CSExCv}.  
As noted in Refs. [3], and confirmed here, the second order phase transition present for 
$\Delta_{\rm mp}$ is considerably altered by fluctuations. Notably,  the $C_V$ vs $T$ curve is devoid of a discontinuity at $T_c$ with smoothly varying gaps.  

\section{The Random Spacing Model}

Recently, the random spacing (RS) model in which the sp energy levels are randomly distributed around the Fermi energy to mimic those of nuclei obtained via the use of different energy density functionals (EDF's) was introduced [18]. 
In a set consisting of a very large number of randomly generated sp levels for a given nucleus, some are likely to represent the true situation especially considering the variation that exists when different  EDF's and pairing schemes are used. 

\begin{figure}
\includegraphics[width=0.5\columnwidth]{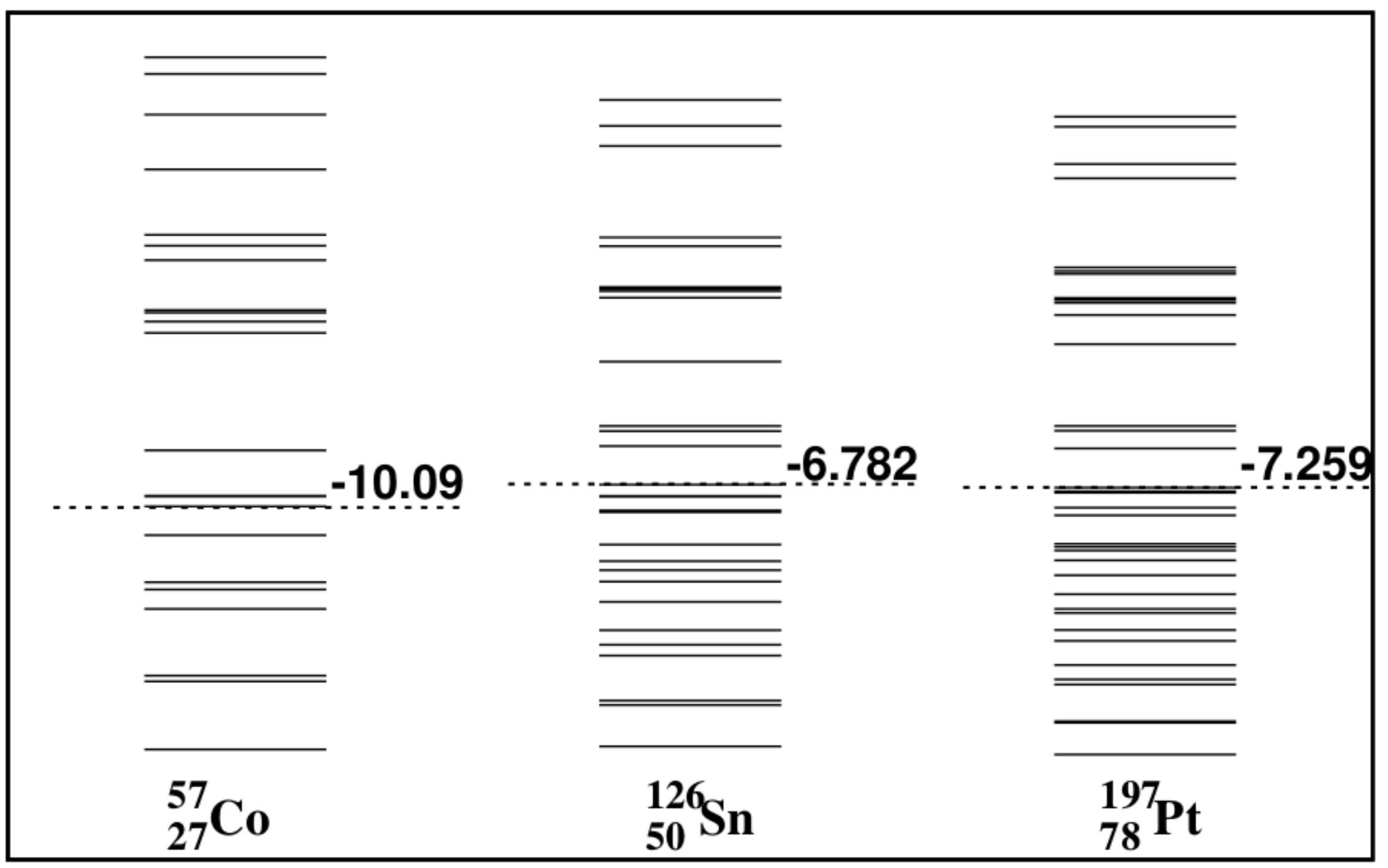}
\includegraphics[width=0.5\columnwidth]{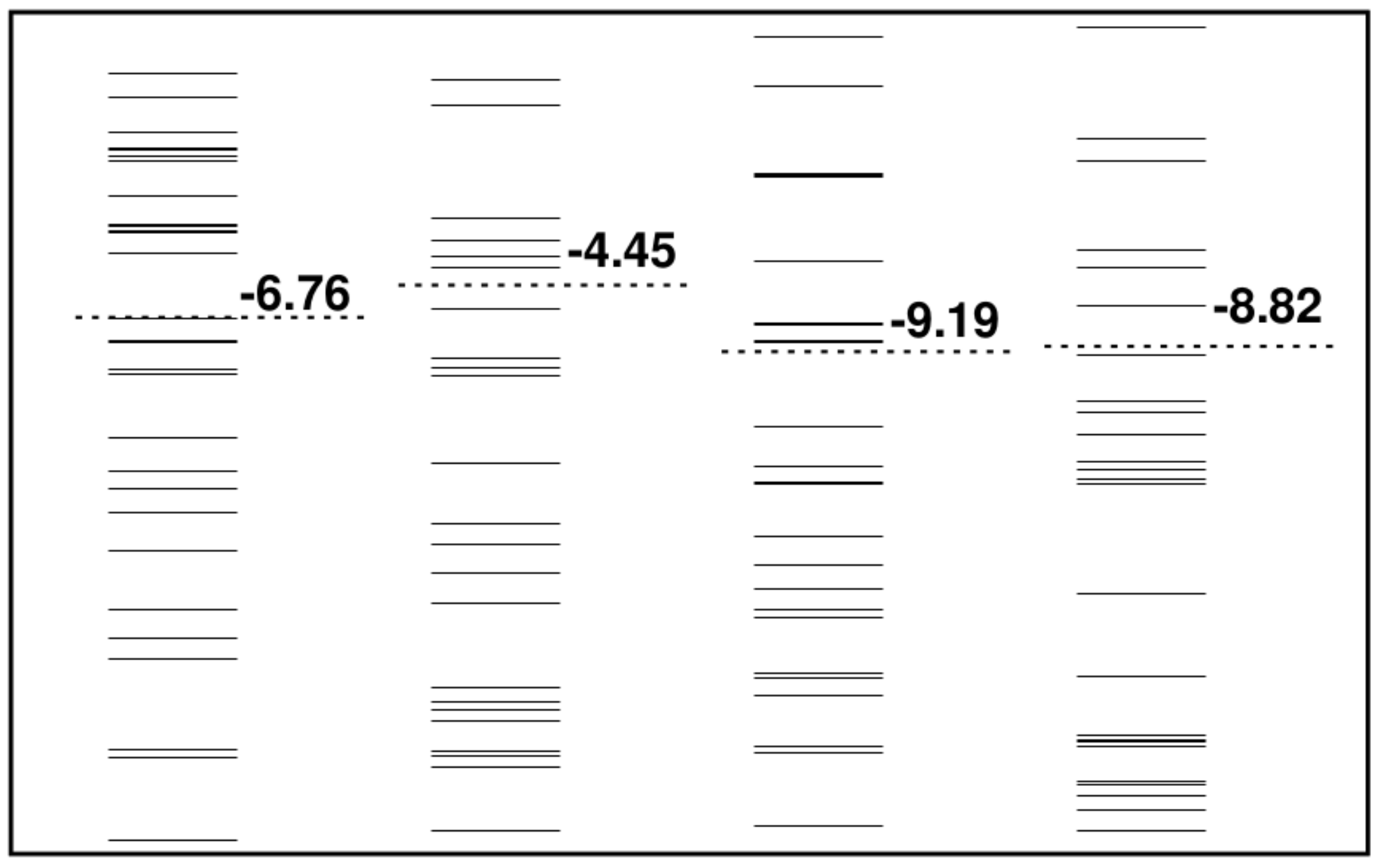}
\caption{(Left) Single particle energy levels of nuclei from HFB calculations using a Skyrme EDF (SkO$^\prime$) [19-21].
(Right) Results of HFB calculations for $N=76$ and three realizations from the RS model. Dotted lines represent the Fermi surface. Figure adapted from [18]. }
\label{RS}
\end{figure}

Figure \ref{RS} presents an illustration of the similarity between the sp energy levels of nuclei from Hartree-Fock-Bogoliubov calculations using a Skyrme EDF (SkO$^\prime$) [19-21] and those of the RS model.  
One advantage of this model is that using easily generated sp levels, statistically based bounds can be placed on the pairing properties of each nucleus.

\begin{figure}
\includegraphics[width=0.5\columnwidth]{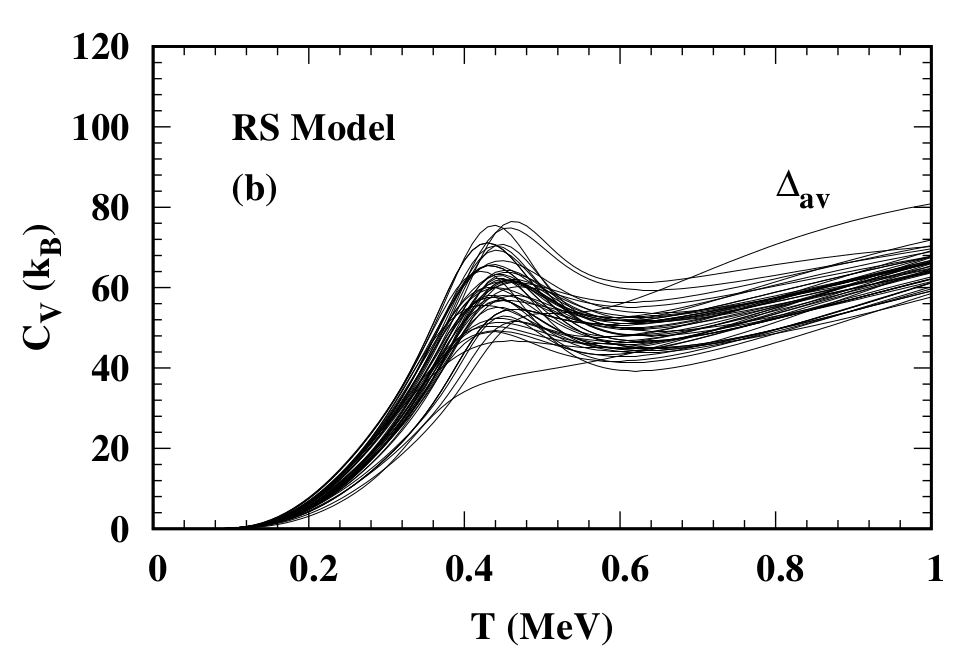}
\includegraphics[width=0.5\columnwidth]{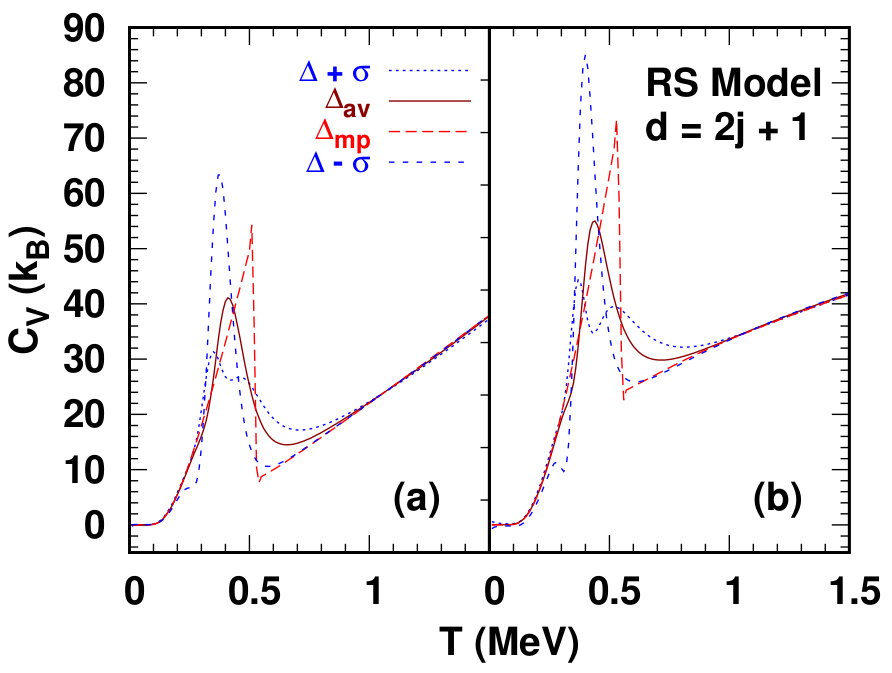}
\caption{Specific heat obtained using the RS model with the inclusion of fluctuations 
(left). Two independent realizations of the RS model with use of the
most probable and average gaps including 1-$\sigma$ deviations (right). Figure adapted from [18]. }
\label{cvrs}
\end{figure}

Including fluctuations using $\Delta_{\rm av}$ as outlined in Sec. 2, the specific heat $C_V$ as a function of temperature is shown 
in  Fig. \ref{cvrs} using  a large number of the RS model sp energy levels. 
The levels were randomly distributed within  a  window of $2\hbar \omega$ around the Fermi level 
for $N=144$. Each level was endowed with the degeneracy $d=2j+1$ characteristic of shell model sp energy levels 
with angular momentum $j$. Increasing the number of random realizations in the ensemble 
makes the band denser, but the borders remain more or less the same. This feature indicates that 
results obtained using  realistic  EDF's would lie within the band shown.   This feature is particularly useful for performing sensitivity tests in astrophysical settings that harbor exotic nuclei. Note also the absence of a second order phase transition as evidenced by the shoulder-like or S-shaped structure of $C_V$ around $T_c$ of the mean field BCS model.

Results of $C_V$ using $\Delta_{\rm mp}$, $\Delta_{\rm av}$ and $\Delta_{\rm av} \pm \sigma$ for two realizations 
among hundreds of individual random realizations of sp energy levels are shown in the right panel of  Fig. \ref{cvrs}. 
Although the overall features in this figure are similar to those of the CS Model, quantitative differences exist 
owing to the different bunching and degeneracy of the individual sp energy levels of the RS model.

\section{Outlook}

Calculations of level densities and the spin distributions of nuclei including fluctuations in the RS model are in progress and will be reported elsewhere. 
A semiclassical treatment of fluctuations is strictly valid only when the mean sp level spacing around the Fermi surface is smaller or nearly equal to the zero temperature pairing gap  and a fully quantum treatment of fluctuations becomes necessary otherwise to overcome the limitations of the mean field BCS formalism  [7-17].  Contrasting the semiclassical and quantum treatments of fluctuations  in the canonical and grand canonical approaches [11,17] as well as 
investigations of fluctuations in highly neutron-rich isotopes with more advanced techniques 
in the context of the RS model are other investigations under study. 

\section{Acknowledgements}

Beneficial communications with P.-G. Reinhardt, Steve Grimes, Alexander Voinov and Tom Massey are gratefully acknowledged. This work was performed with research support from the U.S. DOE grant No. DE-FG02-93ER-40756.

%

\end{document}